\documentclass[prd,12pt]{revtex4-1}

\usepackage{amsmath}
\usepackage{commath}
\usepackage{graphicx}
\usepackage{amssymb}
\usepackage{subfigure}
\usepackage{bm}
\usepackage{slashed}
\usepackage{tikz}
\usepackage{braket}
\usepackage[labelsep=space]{caption}
\usepackage[capitalise]{cleveref}

\def\be{\begin{equation}}
\def\ee{\end{equation}}	
\def\bea{\begin{eqnarray}}
\def\eea{\end{eqnarray}}

\def\be{\begin{equation}}
\def\ee{\end{equation}}	
\def\bea{\begin{eqnarray}}
\def\eea{\end{eqnarray}}

\usepackage{slashed}
\usepackage{subfigure}
\usepackage{notoccite}

\def\be{\begin{equation}}
\def\ee{\end{equation}}
\def\bea{\begin{eqnarray}}
\def\eea{\end{eqnarray}}

\def\ba{\begin{array}}
	\def\ea{\end{array}}

\usepackage{amsmath} 
\begin{document}

\title{Diffractive $\phi$  electroproduction with a holographic meson wavefunction\footnote{Contribution to the proceedings of XXIV International Workshop on Deep-Inelastic Scattering and Related Subjects, 11-15 April, 2016, DESY Hamburg, Germany}} 


\author{Mohammad Ahmady}
\email{mahmady@mta.ca}
\affiliation{\small{Department of Physics, Mount Allison University, Sackville, New Brunswick, Canada E4L 1E6} }
	
\author{Ruben Sandapen }
\email{ruben.sandapen@acadiau.ca}
\affiliation{\small{Department of Physics, Acadia University,
	Wolfville, Nova-Scotia, Canada, B4P 2R6 and
	Department of Physics, Mount Allison University,
	Sackville, New Brunswick, Canada, E4L 1E6} }

\author{Neetika Sharma}
\email{neetika@iisermohali.ac.in}
\affiliation{\small{Department of Physical Sciences,
	Indian Institute of Science Education and Research Mohali,
	S.A.S. Nagar, Mohali-140306, Punjab, India} }

	\begin{abstract}
		We predict the cross-section for diffractive $\phi$ electroproduction within the dipole model, using a holographic meson wavefunction for the $\phi$. For the dipole cross-section, we use the Color Glass Condensate dipole model whose parameters are fitted to the latest 2015 combined HERA data on Deep Inelastic Scattering. Choosing a  strange quark mass of $0.14$ GeV, we find good agreement with the available data.
	\end{abstract}
	
	\maketitle
 
\section{Introduction}
\label{Introduction}

Light-front holography, pioneered by Brodsky and de T\'eramond, is a remarkable correspondance between a semiclassical  approximation (massless quarks and no quantum loops) of light-front QCD and gravity in the higher dimensional anti-de Sitter (AdS) space. For an excellent  review of this approach, we refer to \cite{Brodsky:2014yha}. Light-front holography provides analytical forms for the non-perturbative meson light-front wavefunctions with a single  parameter, a mass scale $\kappa$.  This mass scale breaks conformal invariance in the infrared region of AdS and in physical spacetime, it is chosen to fit the observed slope of meson Regge trajectories.  For example, Ref. \cite{Brodsky:2014yha} reports $\kappa=0.54$ GeV for vector mesons. With $\kappa$ thus fixed, light-front holography predicts the same light-front wavefunction for the $\rho$ and $\phi$ vector mesons. To distinguish between these two vector mesons, one has to account for non-zero quark masses and this can be done, at least for light quark masses, following the prescription of Brodsky and de T\'eramond \cite{Brodsky:2008pg}. We follow this prescription here and  we treat the quark mass as a free parameter. In this contribution, we shall show results using a strange quark mass of $0.14$ GeV as this leads to good agreement with the available data on diffractive $\phi$ meson production.

In Ref. \cite{Forshaw:2012im}, successful dipole model predictions were obtained for diffractive $\rho$ production using the holographic wavefunction for the $\rho$ and the Color Glass Condensate (CGC) dipole cross-section \cite{Iancu:2003ge,Watt:2007nr} whose parameters were fitted to early HERA DIS data. In 2015, the latest high precision combined HERA DIS data were released \cite{Abramowicz:2015mha}.  We use these new data here to update the parameters of the CGC dipole cross-section and make predictions for diffractive $\phi$ production, thereby testing the holographic wavefunction for the heavier $\phi$ meson.  Note that, although we treat the quark mass as a free parameter, we still use the same quark mass when fitting the dipole cross-section to DIS data and when making predictions for vector meson production. 

In what follows, we shall focus on the features of the holographic meson light-front wavefunction and then report our predictions for diffractive $\phi$ meson production. Further details can be found in \cite{Ahmady:2016ujw}. 

\section{Holographic light-front meson wavefunction}
The dipole model of high energy scattering is well-known and highly successful. The largeness of the center-of-mass energy, $\sqrt{s}$, guarantees the factorization of the imaginary part of the scattering amplitude. For diffractive vector meson production, in a standard notation, the imaginary part of the forward scattering amplitude is given by
\bea
\Im \mbox{m}\, \mathcal{A}_\lambda(s;Q^2)  
&=&   s \sum_{h, \bar{h}} \int {\mathrm d}^2 {\mathbf r} \; {\mathrm d} x \; \Psi^{\gamma^*,\lambda}_{h, \bar{h}}(r,x;Q^2)  \Psi^{V,\lambda}_{h, \bar{h}}(r,x)^*  \sigma(x_{\text{m}},\mathrm{r})
\label{amplitude-VMP} 
\eea
where $\Psi^{\gamma^*,\lambda}_{h, \bar{h}}(r,x;Q^2)$ and $\Psi^{V,\lambda}_{h, \bar{h}}(r,x)$  are the light-front wavefunctions of photon and vector meson respectively while $\sigma (x_{\text{m}},\mathrm{r})$ is the dipole cross-section. It is worth noting that the high energy factorization in equation \eqref{amplitude-VMP} is independent of the validity of perturbation theory, i.e. it is valid for all dipole sizes. In practice, the photon light-front wavefunctions and dipole-proton amplitude are computed using perturbative QED and high energy perturbative QCD respectively. Note that the quark mass acts as an infrared regulator in the perturbative expressions for the photon light-front wavefunctions. On the other hand, perturbation theory is useless to find the meson wavefunction and here we use light-front holography to predict it. 

  In a semiclassical approximation of light-front QCD with massless (and spinless) quarks, the meson wavefunction can be written as \cite{Brodsky:2014yha} 

\begin{equation}
	\Psi(\zeta, x, \phi)= e^{iL\phi} \mathcal{X}(x) \frac{\phi (\zeta)}{\sqrt{2 \pi \zeta}} 
	\label{mesonwf}
\end{equation}
where the variable $\zeta=\sqrt{x(1-x)} r$ is the transverse separation between the quark and the antiquark at equal light-front time. The transverse wavefunction $\phi(\zeta)$ is a solution of the so-called holographic light-front Schr\"odinger equation:
\begin{equation}
	\left(-\frac{d^2}{d\zeta^2} - \frac{1-4L^2}{4\zeta^2} + U(\zeta) \right) \phi(\zeta) = M^2 \phi(\zeta) 
	\label{holograhicSE}
\end{equation}
where $M$ is the mass of the meson and $U(\zeta)$ is the confining potential which at present cannot be computed from first-principle in QCD. On the other hand, making the substitutions $\zeta \to z$ where $z$ being the fifth dimension of AdS space, together with   $L^2 -(2-J)^2 \to (mR)^2$  where $R$ and $m$ are the radius of curvature and mass parameter of AdS space respectively, then Eq. \eqref{holograhicSE} describes the propagation of spin-$J$ string modes in AdS space. In this case, the potential is given by
\begin{equation}
	U(z, J)= \frac{1}{2} \varphi^{\prime\prime}(z) + \frac{1}{4} \varphi^{\prime}(z)^2 + \left(\frac{2J-3}{4 z} \right)\varphi^{\prime} (z) 
\end{equation}
where $\varphi(z)$ is the dilaton field which breaks the conformal invariance of AdS space. A quadratic dilaton ($\varphi(z)=\kappa^2 z^2$) profile results in a harmonic oscillator potential in physical spacetime:
\begin{equation}
	U(\zeta,J)= \kappa^4 \zeta^2 + \kappa^2 (J-1) \;.
	\label{harmonic-LF}
\end{equation}
Brodsky, Dosch and de T\'eramond have shown  that the light-front harmonic potential is unique \cite{Brodsky:2013ar}.  Solving the holographic Schr\"odinger equation with this harmonic potential given by Eq. \eqref{harmonic-LF} yields the meson mass spectrum,
\begin{equation}
	M^2= 4\kappa^2 \left(n+L +\frac{S}{2}\right)\;
	\label{mass-Regge}
\end{equation}
with the corresponding normalized eigenfunctions
\begin{equation}
	\phi_{n,L}(\zeta)= \kappa^{1+L} \sqrt{\frac{2 n !}{(n+L)!}} \zeta^{1/2+L} \exp{\left(\frac{-\kappa^2 \zeta^2}{2}\right)} L_n^L(x^2 \zeta^2) \;.
	\label{phi-zeta}
\end{equation}
To completely specify the holographic wavefunction given by Eq. \eqref{mesonwf}, the longitudinal wavefunction $\mathcal{X}(x)$ must be determined. For massless quarks, this is achieved by an exact mapping of the pion  electromagnetic form factors in AdS and in physical spacetime resulting in  \cite{Brodsky:2014yha}
\begin{equation}
	\mathcal{X}(x)=\sqrt{x(1-x)}
\end{equation}

It remains to restore the helicity and quark mass dependence of the meson light-front wavefunctions. This results in \cite{Forshaw:2003ki,Forshaw:2012im}
\be
\Psi^{V,L}_{h,\bar{h}}(r,x) =  \frac{1}{2} \delta_{h,-\bar{h}}  \bigg[ 1 + 
{ m_{f}^{2} -  \nabla_r^{2}  \over x(1-x)M^2_{V} } \bigg] \Psi_L(r,x) 
\label{mesonL}
\ee
and
\be \Psi^{V, T}_{h,\bar{h}}(r,x) = \pm \bigg[  i e^{\pm i\theta_{r}}  ( x \delta_{h\pm,\bar{h}\mp} - (1-x)  \delta_{h\mp,\bar{h}\pm})  \partial_{r}+ m_{f}\delta_{h\pm,\bar{h}\pm} \bigg] {\Psi_T(r,x) \over 2 x (1-x)} 
\label{mesonT}
\ee
where the subscripts $L,T$ on $\Psi(r,x)$ indicate that we choose the normalization of the light-front wavefunction to be polarization dependent, i.e. we use the normalization condition \cite{Forshaw:2012im}
\be
\sum_{h,\bar{h}} \int {\mathrm d}^2 {\mathbf{r}} \, {\mathrm d} x |
\Psi^{V, \lambda} _{h, {\bar h}}(x, r)|^{2} = 1 \;.
\ee
Having specified the meson light-front wavefunctions, we are now able to compute the cross-section for diffractive $\phi$ meson production.
\section{Results}

We compute the  differential cross-section in the forward limit, i.e.

\be
\left. {{\mathrm d} \sigma_\lambda \over dt}\right. \mid_{t=0}  
= \frac{1}{16\pi} 
[ \mathcal{A}_\lambda(s, t=0)]^2
\label{dxsection}
\ee
where $\mathcal{A}_{\lambda}$ includes the (small) real part correction of the amplitude.  We assume the $t$-dependence to be exponential, i.e.   
\be
\left. {{\mathrm d} \sigma_\lambda \over dt}\right.
= \frac{1}{16\pi} 
[\mathcal{A}_\lambda(s, t=0)]^2  \exp(-B_D t)
\label{dsigmadt}
\ee
where the diffractive slope parameter $B_D$  is given by
\be
B_D = N\left( 14.0 \left(\frac{1~\mathrm{GeV}^2}{Q^2 + M_{V}^2}\right)^{0.2}+1\right)
\label{Bslope}
\ee
with $N=0.55$ GeV$^{-2}$. 

\begin{figure}[htbp]
	\centering 
	\includegraphics[width=.60\textwidth]{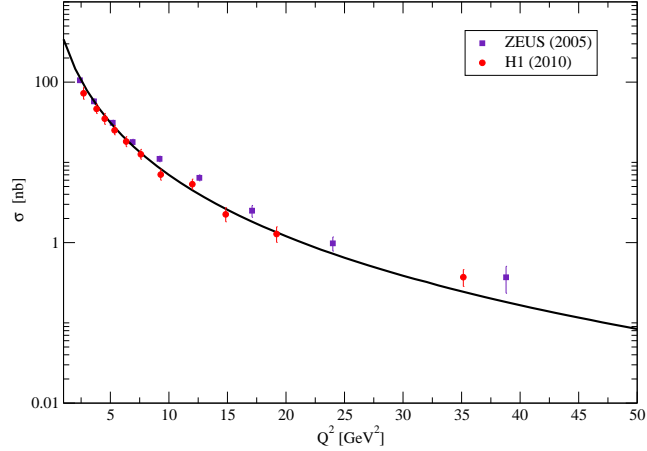}
	\caption{ Predictions for the $\phi$ production total cross section for at $W=90$ GeV as a function of $Q^2$  compared to HERA data \cite{Chekanov:2005cqa,Aaron:2009xp}. The theory prediction is generated using a strange quark mass, $m_s=0.14$ GeV.}
	\label{phi-q2}
\end{figure}

\begin{figure}[htbp]
	\centering 
	\includegraphics[trim={0 9.5cm 0 3.5cm},width=.70\textwidth]{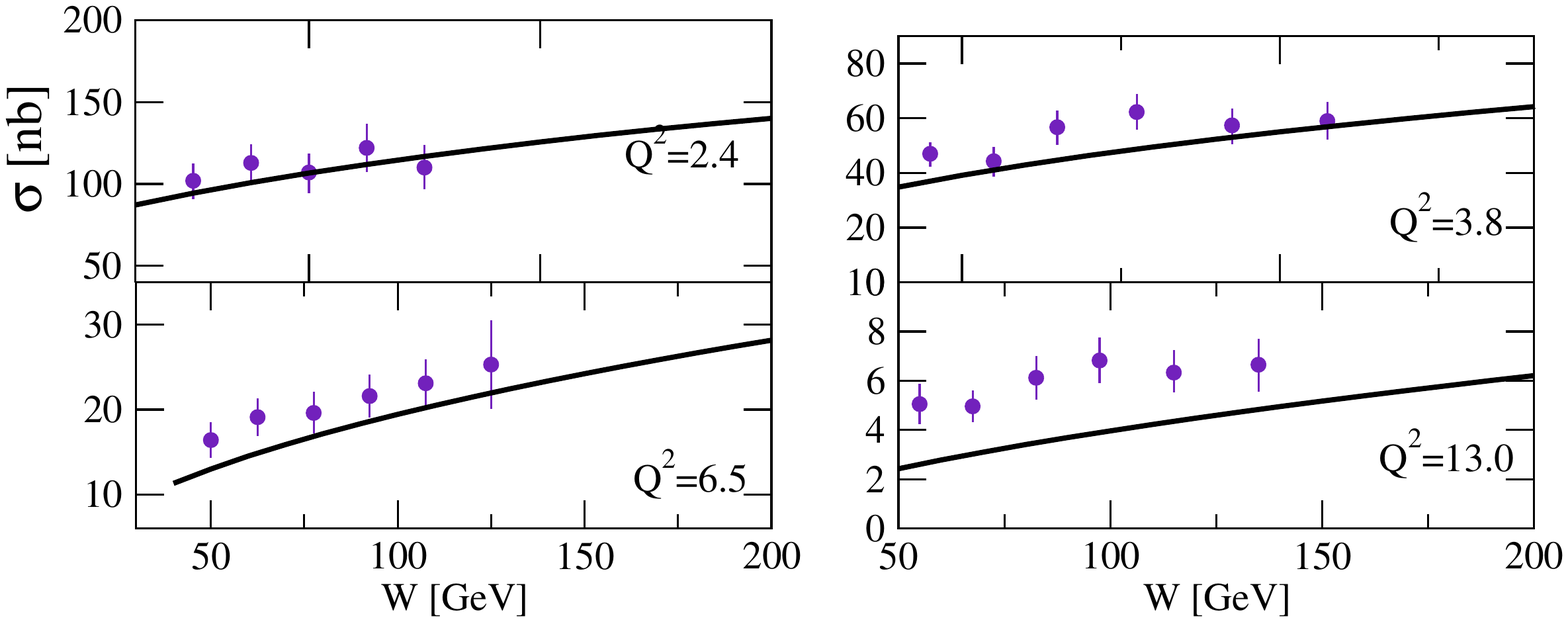}
	\caption{Our predictions for $\phi$ production total cross-section for $\gamma^* p \to \phi p$ as a function of $W$ in different $Q^2$ bins compared to the ZEUS data \cite{Chekanov:2005cqa}. The theory prediction is generated using a strange quark mass, $m_s=0.14$ GeV.} 
	\label{phi-W-ZEUS}
\end{figure}

\begin{figure}[htbp]
	\centering 
	\includegraphics[trim={0 9.5cm 0 2cm},width=.70\textwidth]{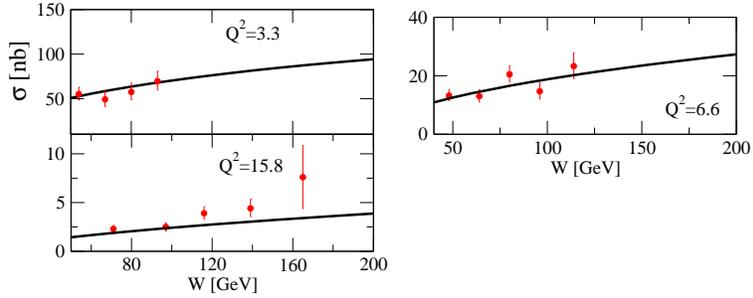}
	\caption{Our predictions for the $\phi$ production total cross-section as a function of $W$ in different $Q^2$ bins compared to the H1 data \cite{Aaron:2009xp}. The theory prediction is generated using a strange quark mass, $m_s=0.14$ GeV.} 
	\label{phi-W-H1}
\end{figure}

\begin{figure}[htbp]
	\centering 
	\includegraphics[trim={0 0 0 2cm},width=.60\textwidth]{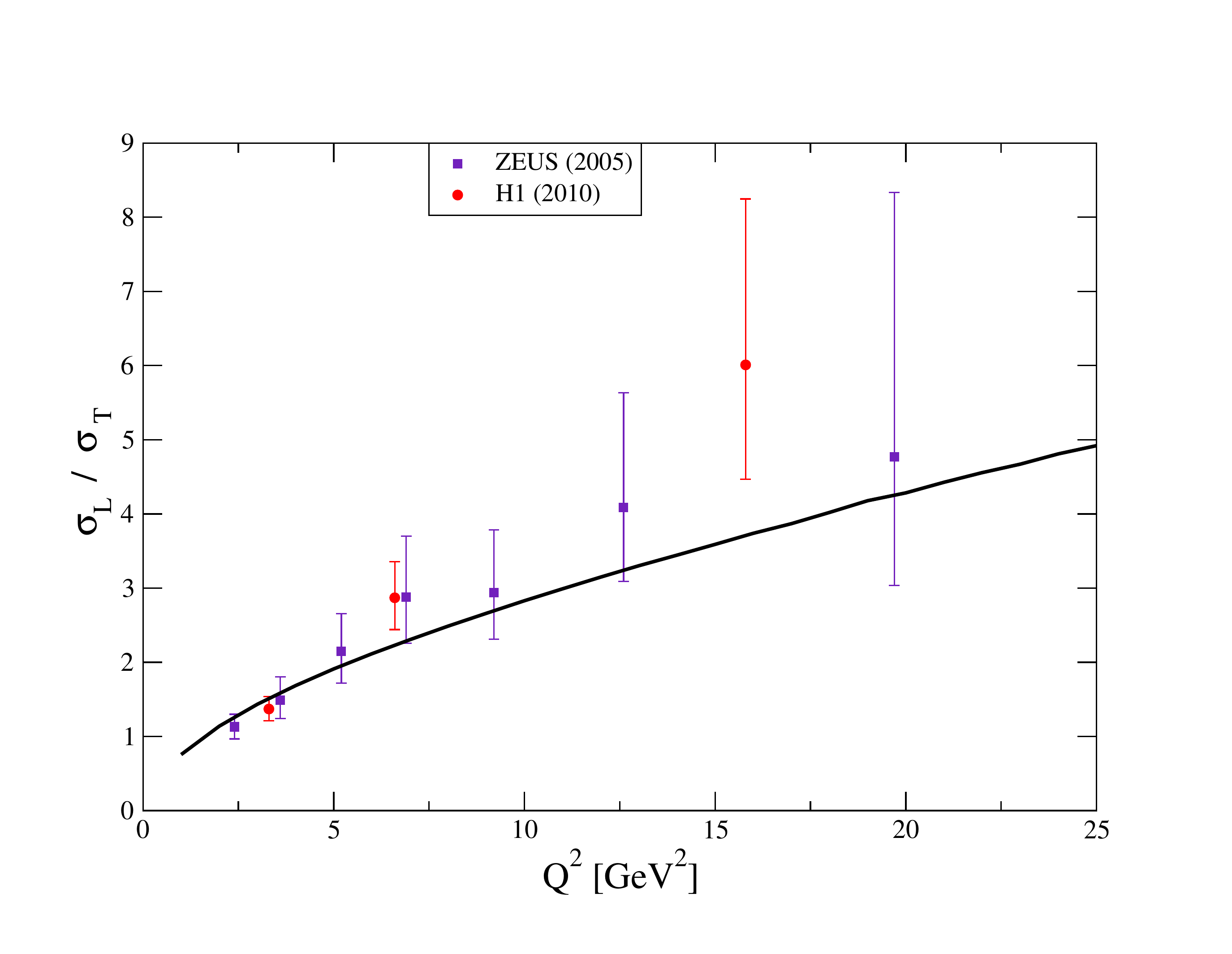}
	\caption{Our predictions for the $\phi$ production longitudinal to transverse cross-section ratio at $W=90$ GeV compared to the ZEUS data \cite{Chekanov:2005cqa}. The theory prediction is generated using a strange quark mass, $m_s=0.14$ GeV.} 
	\label{phi-ratio}
\end{figure}

As can be seen from figures \ref{phi-q2},\ref{phi-W-ZEUS},\ref{phi-W-H1} and \ref{phi-ratio}, agreement with data is good except at the highest $Q^2$ (especially for the ZEUS data at $Q^2=13~\text{GeV}^2$). This is presumably due to the fact that the holographic wavefunction neglects the perturbative evolution of the meson wavefunction.

\section{Conclusion}
We have used the light-front wavefunction for the $\phi$ meson as predicted by light-front holography to generate predictions for the cross-section of diffractive $\phi$ meson production within the dipole model. Agreement with the available HERA data is good and could possibly be improved at high $Q^2$ if the perturbative evolution of the holographic light-front meson wavefunction is taken into account.

\section{Acknowledgements}
This work of NS is supported by the Department of Science and Technology, Government of India, under the Fast Track scheme (Ref No. SR/FTP/PS-057/2012). The work of MA and RS is supported by a team grant (SAPGP-2014-00002) from the National Science and Engineering Research Council of Canada (NSERC).
 

\bibliographystyle{JHEP}
\bibliography{DIS2016}

\end{document}